\documentclass{article}

\usepackage{amsmath}
\usepackage{arxiv}

\usepackage[utf8]{inputenc} 
\usepackage[T1]{fontenc}    
\usepackage{hyperref}       
\usepackage{url}            
\usepackage{booktabs}       
\usepackage{amsfonts}       
\usepackage{nicefrac}       
\usepackage{microtype}      
\usepackage{lipsum}		
\usepackage{graphicx}
\usepackage{natbib}
\usepackage{doi}
\usepackage{siunitx}
\usepackage[outdir=./]{epstopdf}

\makeatletter
\providecommand\add@text{}
\newcommand\tagaddtext[1]{%
  \gdef\add@text{#1\gdef\add@text{}}}%
\renewcommand\tagform@[1]{%
  \maketag@@@{\llap{\add@text\quad}(\ignorespaces#1\unskip\@@italiccorr)}%
}
\makeatother

\newcommand{\orc}{\raisebox{-.2\height}{\includegraphics[width=10pt]{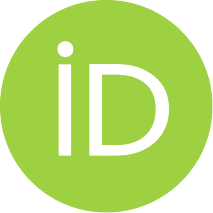}}}

\title{Injection rate of cylinder lubrication oil in large two-stroke marine diesel engines using a common rail lubrication system}

\author{ \href{https://orcid.org/0000-0001-8835-4038}{\includegraphics[scale=0.06]{orcid.pdf}\hspace{1mm}Bj{\o}rn Christian Dueholm}
    \\
	Department of Materials and Production\\
	Aalborg University\\
	Fredrik Bajers Vej 7K, 9220 Aalborg East, Denmark\\
	\And
	\href{https://orcid.org/0000-0001-5501-7019}{\includegraphics[scale=0.06]{orcid.pdf}\hspace{1mm}Jesper de Claville Christiansen} \\
	Department of Materials and Production\\
	Aalborg University\\
	Fredrik Bajers Vej 7K, 9220 Aalborg East, Denmark\\
 	\And
	\href{https://orcid.org/0000-0002-7013-1490}{\includegraphics[scale=0.06]{orcid.pdf}\hspace{1mm}Benny Endelt} \\
	Department of Materials and Production\\
	Aalborg University\\
	Fredrik Bajers Vej 7K, 9220 Aalborg East, Denmark\\
  	\And
	Nikolaj Kristensen \\
	Hans Jensen Lubricators A/S\\
	Hadsund, Denmark\\
}

\renewcommand{\shorttitle}{ROI of cylinder lubrication oil in large two-stroke marine diesel engines}

\hypersetup{
pdftitle={A template for the arxiv style},
pdfsubject={q-bio.NC, q-bio.QM},
pdfauthor={David S.~Hippocampus, Elias D.~Striatum},
pdfkeywords={First keyword, Second keyword, More},
}

\begin{document}
\maketitle

\begin{abstract}
This paper investigates a common rail cylinder lubrication system for large two-stroke marine diesel engines using electronically controlled injectors. The system is studied using the Bosch rate of injection measurement technique The common rail injector has a buildup of mass flow of approximately 1  \si{\milli\second} as the injector opens until the nozzle is choked from cavitation. Using a highly viscous fluid, the Bosch rate of injection method is able to predict the injected amount with an error of 5\% or lower for nearly the entire tested delivery range of 2 \si{\milli\gram} to 21 \si{\milli\gram}. Lubrication of cylinder liners and piston rings is a crucial parameter in operating a two-stoke marine diesel engine efficiently. Both over and under lubrication is harmful for the engine, so the ability to accurately dose the cylinder oil is very important. A mass flow build up time of 1 ms promises high accuracy of dosage even down to 2.5 \si{\milli\gram} per injection. This paves the way for injecting the oil where and when it is needed, which in turn will improve engine performance and lower harmful emissions.
\end{abstract}


\section{Introduction}
The slow speed uniflow scavenged two-stroke engine has become the most widespread propulsion system for large-scale ships \cite{2009_WOODYARD}. With the continuous developments in both the construction and operation of the engines, new challenges have arisen. Recently, the environment and the emission of harmful particles have been in focus. The regulations for where and how much vessels are allowed to emit are becoming stricter and this forces the industry to continue evolving to meet new standards. These regulations include both the elimination of greenhouse gasses from shipping by the end of the century and a reduction in harmful emissions such as SOx, NOx and particulate matter \cite{2018_IMO}. Cylinder Lubrication oil contributes to the emission of particulate matter when it is left on the cylinder liner exposed to the combustion taking place in the engine. Since it is necessary to have the oil on the cylinder liner, and exposing it to the combustion is unavoidable, only using the necessary amount is key to reducing the amount lost to combustion. The contribution to potential particulate matter emissions from ships is approximately divided evenly between lubrication oil and fuel \cite{2004_Dragsted,Aabo2001}.\\
With lubrication oil playing a large role in emissions from vessels, the recent focus on the climate and the fact that there is a potential for saving money, optimising lubrication is a logical step. Optimising lubrication with regards to using a smaller amount of lubrication oil will reduce the emission of particulate matter and cut operational costs.\\
Since there is evidently an incentive to optimise the lubrication, other authors have also looked into the topic. The other investigations have primarily been in the following areas: optimising control of the lubrication system, interaction of the piston ring and the oil film, and designing a new lubrication system.
One author explored the topic of optimising control of the injection of lubrication oil on an electronically controlled cylinder lubrication system (ECCLS) \cite{2019_Yuhai, He2019}. The ECCLS investigated is similar to the electronically controlled systems produced by many engine manufacturers. The system injects cylinder lubrication oil into the piston ring pack, however, there are however some differences with regards to the control algorithm. The system was tested on a real engine and showed improved cylinder condition and resulted in 24.65\% in lubrication oil savings.\\
Several authors have studied the interaction between the piston ring pack and liner surface related to marine lubrication \cite{Li2022,Li2021,Overgaard2018,Xiuyi2022,Jiao2021}. The studies employ the average Reynolds equation to mathematically model the sliding of a ring on the liner surface. The focus of these papers was to enhance the lubrication to reduce liner wear and improve gas sealing. These studies all assume that there is a well distributed oil film on the liner surface. Many conventional systems use non return valves for delivering the oil in the piston ring pack, as the piston passes the injection point. The oil is then distributed by the ring pack onto the liner surface \cite{2019_Yuhai}. \\
Recently authors have investigated designing new systems for the lubrication delivery \cite{Milanese2021}. The performance of the injectors was simulated using AMESim, and they found using a common rail system to inject into the piston ring pack led to over 50\% reduction in loss of oil measured in the exhaust using a sulphur tracing technology. Many of the advantages of the common rail systems are already known from the fuel delivery systems on the same engines. More direct pressure buildup leads to a more instantaneous injection, which makes it easier to control the delivered amount over the load range of the engine.\\

These studies all try to optimise delivered amount, distribution of the oil or interaction between rings and liner assuming an even distribution. Based on this, it is clear that there is a need for a deeper understanding of the injection profile of the cylinder lubrication oil injectors to better determine the timing and distribution of the injected cylinder oil. This was also the case for the fuel delivery systems on the same engines, the injection rates of which were also investigated \cite{Payri2021} as a natural step in optimising timing and distribution. To determine the rate of injection, the Bosch rate of injection method is used, which was designed to characterise fuel injectors and was therefore designed to work with low viscous fluids. Because of this, the aim of this study is twofold:
\begin{itemize}
    \item Experimentally characterising an electronically controlled valve injecting marine cylinder lubrication oil.
    \item Verifying the validity of the Bosch rate of injection method for highly viscous fluids.
\end{itemize}
The cylinder lubrication system (CLS) used to investigate this is the only common rail CLS on the market for two-stroke marine diesel engines. The system uses single hole injectors without a return flow pipe. The valves are electronically controlled by a solenoid and spray the oil in an upwards motion towards the top of the liner where the wear is highest and the demand for lubrication oil largest. The injection of oil happens before the piston passes the injection points and relies on the swirling motion of the scavenging air to deposit the oil on the liner surface. The technology is known as the swirl injection principle (SIP), and when using this technology, the rate of injection becomes very interesting, as it strongly influences how the system should be timed and the breakup of the oil, again quite similar to fuel injection systems \cite{Payri2021}.
\section{Materials and Methods}
This section presents the measurement setup and the methodology used for processing and understanding the data.
\subsection{Experimental setup}
\begin{figure}[htb!]
    \centering
    \includegraphics[width=0.5\linewidth,bb=0 0 410 850]{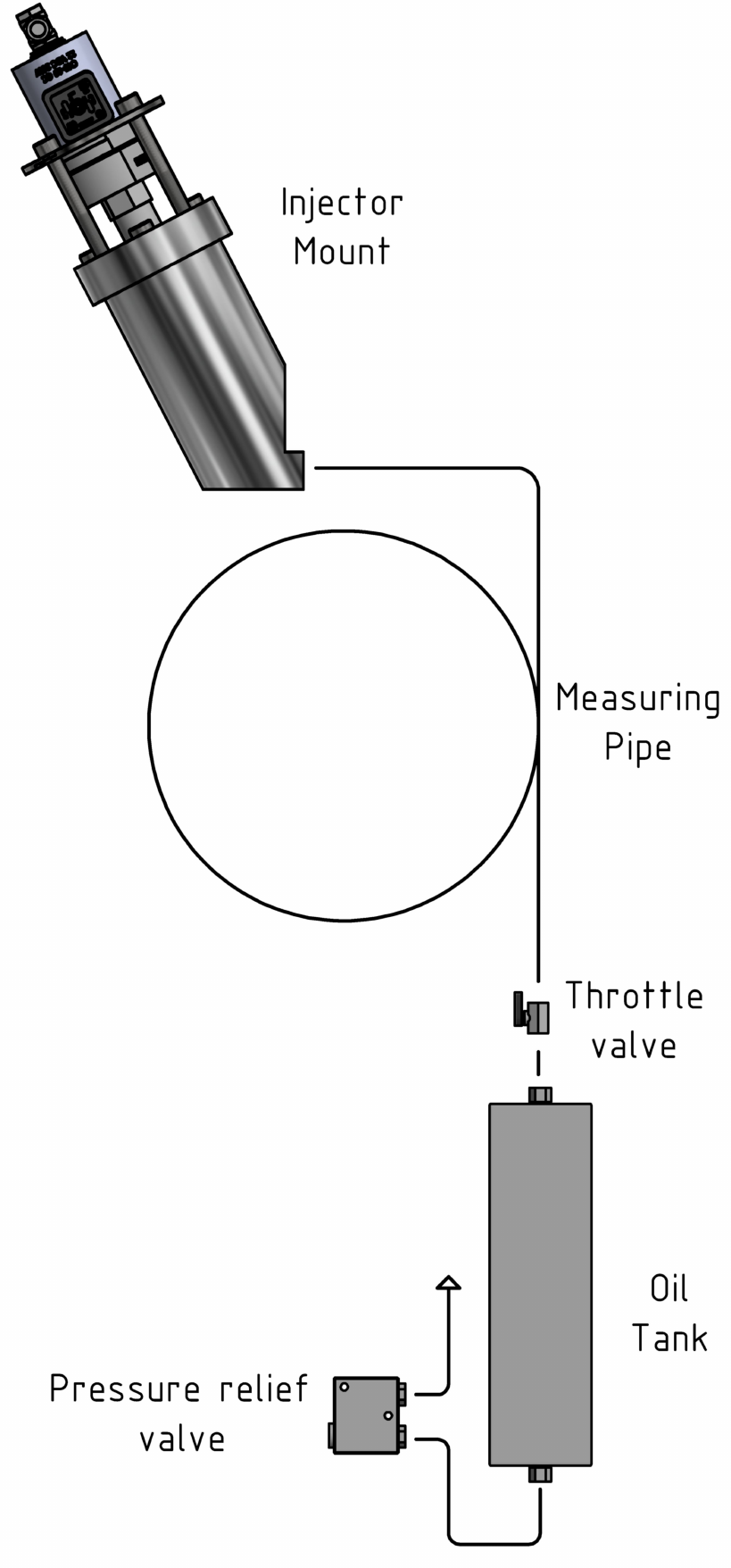}
    \caption{Bosch rate of injection experimental setup used.}
    \label{fig:ROI_setup}
\end{figure}

The principle of measurement is that of \cite{1966_Bosch}. It is called the Bosch rate of injection (ROI) principle and relies on a propagating pressure wave. The setup of the experiment is shown in Figure \ref{fig:ROI_setup}.

\begin{figure}[!hb]
    \centering
    \includegraphics[]{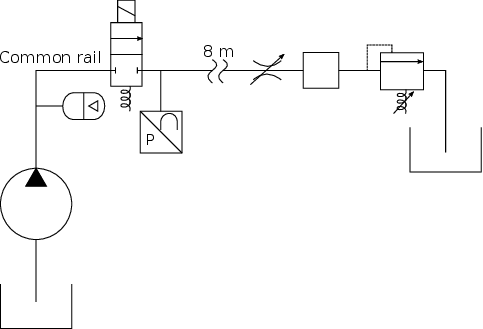}
    \caption{Principal diagram of the experimental setup.}
    \label{fig:ROI_setup_hydraulic}
\end{figure}

The system consists of a pump that pressurises a common rail, an injector which is connected to the common rail, a pressure sensor, a long pipe, a throttle valve, an oil tank and a pressure relief valve. The overall hydraulic overview of the system is presented in Figure \ref{fig:ROI_setup_hydraulic}. The injector is a single hole solenoid actuated valve, which injects oil into the measuring pipe when a voltage is applied to the solenoid. The pressure is measured in the measuring pipe by a piezoelectric pressure sensor. After the pressure sensor is a throttle valve, a tank and a pressure relief valve to maintain a back pressure.

\begin{figure}[htb!]
    \centering
    \includegraphics[width=0.5\linewidth,bb=0 0 610 800]{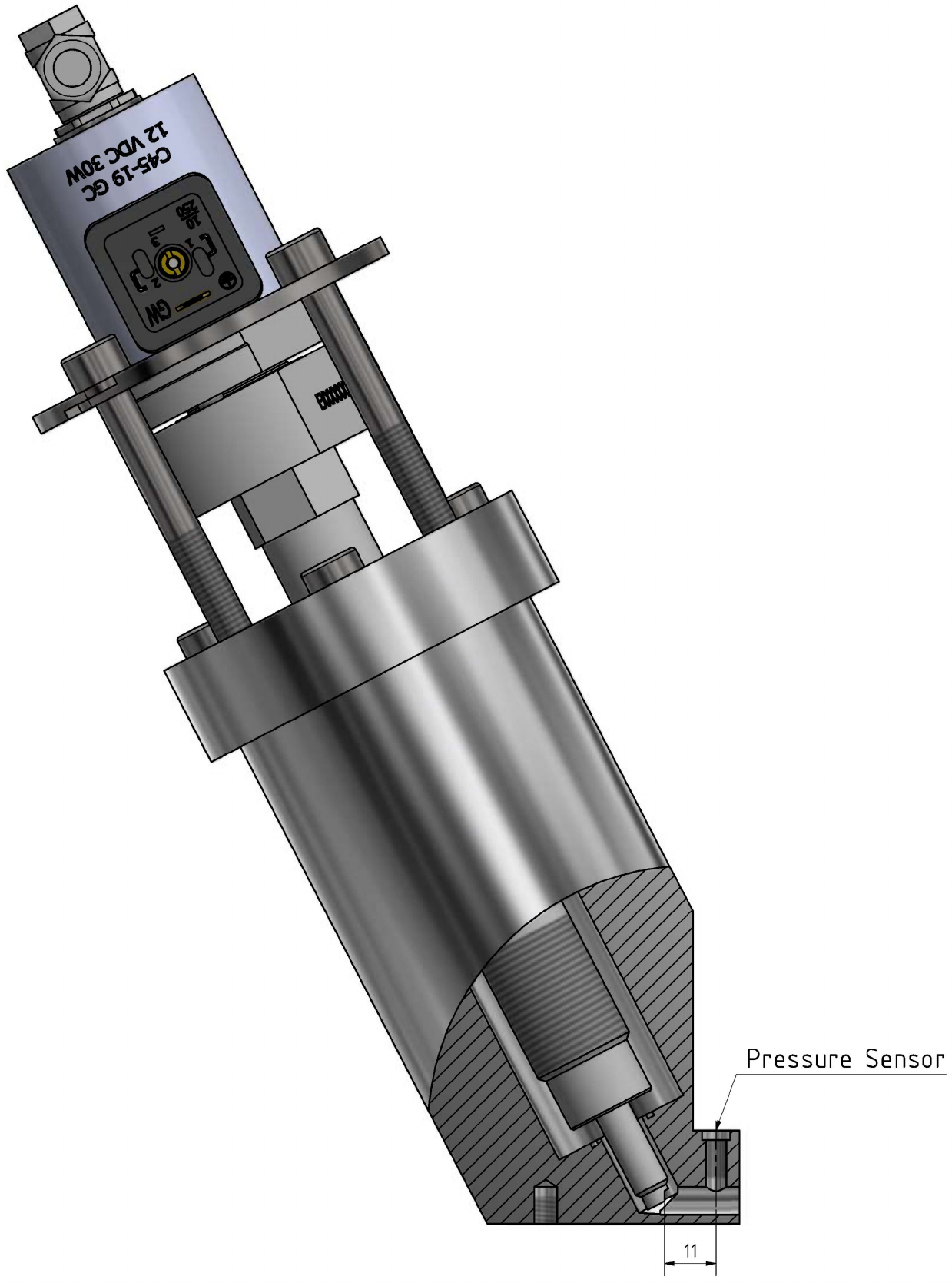}
    \caption{Close-up of the injector mount showing the distance from the tip of the nozzle to the pressure sensor of 11 mm.}
    \label{fig:ROI_mount}
\end{figure}

In Figure \ref{fig:ROI_setup}, the nozzle injects oil into the measuring pipe while being locked in place by the injector mount. The injection of oil raises the pressure locally at the tip of the nozzle and when the pressure equalises with the rest of the pipe, it sends a pressure wave propagating down the measuring pipe. When this pressure wave hits the throttle valve, which is a very slightly open ball valve, it is mostly reflected and travels back along the measuring pipe, being weaker in magnitude the second time it passes the pressure sensor. The pressure sensor measures the pressure in the pipe and is located 11 mm from the tip of the nozzle, as shown in Figure \ref{fig:ROI_mount}.\\

To simulate engine-like conditions the pressure in the pipe is kept at 8 bar by the pressure relief valve at the end of the tank. This is to keep the force balance of the nozzle resembling what it would be were it mounted on an engine.

\begin{figure}
    \centering
    \includegraphics{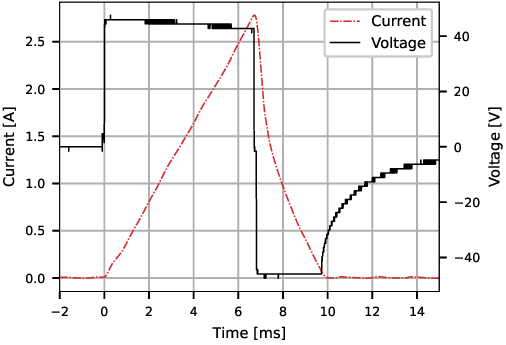}
    \caption{Voltage and current profile used for the injections. Ramp time 6.7 ms.}
    \label{fig:ramping}
\end{figure}

The information for the experimental setup is summarised in Table \ref{tab:exp}.
\begin{table}[hb!]
    
    \centering    
    \caption{Information on the physical geometry and apparatus}
    \begin{tabular}{l l }
        \hline
        Parameter & Value\\
        \hline
        Measuring pipe length & 8 m\\
        Measuring pipe inner diameter & 4 mm\\
        Measuring pipe outer diameter & 6 mm\\
        Tank volume & $1.45\cdot10^{-3}$ m³\\
        Back pressure & 8 bar\\
        Injection pressure & 70 bar\\
        Injection frequency & 1.667 Hz\\
        Working fluid & HydraWay HVXA15 \\
        Pressure sensor & Piezoelectric  \\
        \hline
    \end{tabular}
    \label{tab:exp}
\end{table}
Hydraway HVXA15 was chosen as a working fluid because the viscosity and density at 293.15 K is within 10\% of a typical cylinder oil at 373.15 K \cite{Ravendran2017}. This makes it possible to work with a fluid at room temperature that has the properties similar to a commercial heated cylinder oil. The throttle valve and all fittings used have the same internal diameter to make sure the pressure wave maintains the same shape. The oil is kept at room temperature and the small amount of heat the solenoid is developing is removed by the oil flowing through it and to natural convection. This maintains the stability of the system's temperature and keeps the piezoelectric pressure sensor from drifting. \\
The actuation strategy of the valve is kept relatively simple for the purpose of this experiment. The voltage is turned on for a set amount of time then reversed and slowly turned back to 0. This is shown in Figure \ref{fig:ramping}.
The time period when the voltage is up is defined as the ramp time, due to the ramping nature of the current.
When the data logging occurred several other injectors connected to the same common rail were turned on to increase the flow in the common rail. This is so that the measurements are taken in conditions resembling those of a system mounted on an engine.
At a given ramp time 96 injections are logged. The next section describes how the raw data is processed.

\subsection{Data processing}
\begin{figure}
   \centering
   \includegraphics{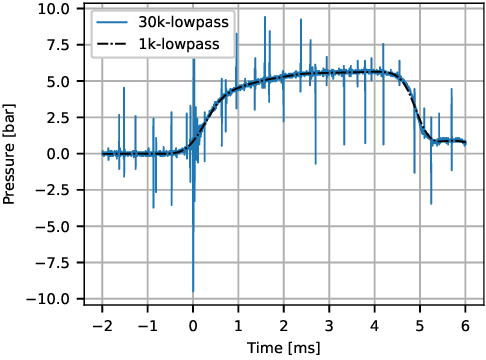}
   \caption{Example of pressure characteristic for 1 kHz and 30 kHz lowpass filters. Ramp time 7.1 ms.}
   \label{fig:raw}
\end{figure}

The pressure variation measured by the pressure sensor in the pipe is used to calculate the volume flow through the nozzle as
\begin{align}
    Q = \frac{\Delta p\cdot A}{\alpha\cdot\rho}\tagaddtext{[m³/s]}
    \label{eq:roi1}
\end{align}
In Equation \ref{eq:roi1}, $Q$ is the volume flow in the pipe in m³/s, $\Delta p$ is the internal pressure fluctuation in Pa, $D$ is the diameter of the pipe in m, $\alpha$ is the speed of sound in the medium in m/s and $\rho$ is the density of the medium in kg/m³. Equation \ref{eq:roi1} is based on the 1-dimensional wave propagation assumption and has been used by authors in the past \cite{1991_Bower,1966_Bosch}.
The data is smoothed with a 1 kHz cutoff lowpass filter before it is used to calculate the rate of injection or the total injection amount. In Figure \ref{fig:raw}, the filtered signal is shown on top of the same signal through a 30 kHz cutoff lowpass filter.
The reason the data is not shown completely unfiltered is because the charge amplifier for the sensor has a lowpass filter installed to protect the equipment. The highest cutoff frequency this lowpass filter can be set to is 30 kHz.

\begin{figure}
    \centering
    \includegraphics{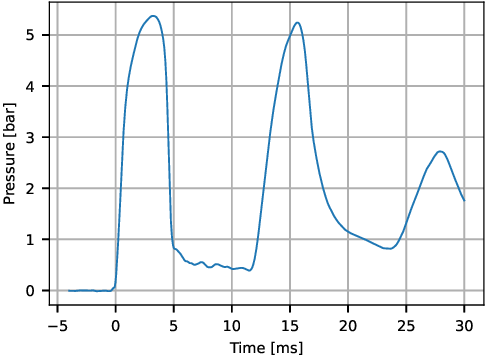}
    \caption{Example of pressure characteristic for a reflected wave. Ramp time 7.1 ms.}
    \label{fig:sos}
\end{figure}

To obtain the speed of sound in the medium, a ball valve is used as a reflector. The ball valve is closed until it is only open very slightly, so effectively working as a throttle. When the pressure wave generated by the nozzle travels along the pipe, it will eventually hit the valve and is reflected, reversing its direction. The wave will then pass the pressure sensor again, where it will be lower in magnitude. This is illustrated in Figure \ref{fig:sos}.
As seen in Figure \ref{fig:sos}, it is quite clear to see when the wave passes the pressure sensor the second time. The distance from the first rising edge to the second is used together with the total length from the pressure sensor to the throttle valve to calculate the speed of sound $\alpha$, which for this setup was measured to be 1375 m/s. A third rising edge can also be seen, which is the second reflection of the wave. The period of the signal is close to constant, which indicates that the wave is not slowing down much but the amplitude is lowered considerably.\\

\begin{figure}
    \centering
    \includegraphics{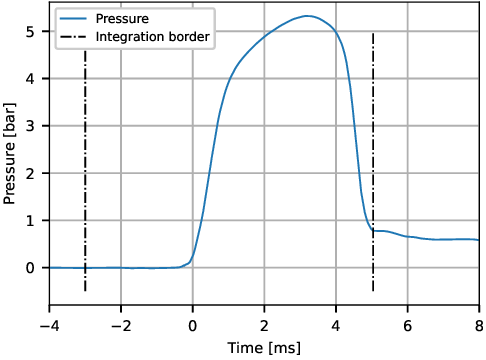}
    \caption{Integration borders of the pressure signal. Ramp time 7.1 ms.}
    \label{fig:integration}
\end{figure}

As the cross-sectional area of the measuring pipe is known together with the speed of sound, Equation \ref{eq:roi1} is used to calculate the mass. The mass is integrated over the injection with the borders of integration, as shown in Figure \ref{fig:integration}, together with a constant injection pressure of 70 bar.
This yields the total mass for the injection, which is used to validate the setup. From Figure \ref{fig:integration}, it is also evident that the rise of mass flow is slowing down in the period from 1 ms to 3 ms. This is due to cavitation in the nozzle, which constricts the flow area and it chokes out the flow of oil through the nozzle. The cavitation in the nozzles is key to getting atomisation in a highly viscous spray system like a common rail lubrication system \cite{Ravendran2019}. The faster cavitation occurs, the faster a spray develops from the nozzle. A faster rise in rate of injection results in a shorter time for spray development, which means less oil is injected as a jet. This is consistent with the goal of the swirl injection principle, which is to atomise the oil and deliver it, with the scavenging air swirl onto the liner surface.
\section{Results and discussion}
\begin{figure}[h!]
    \centering
    \includegraphics{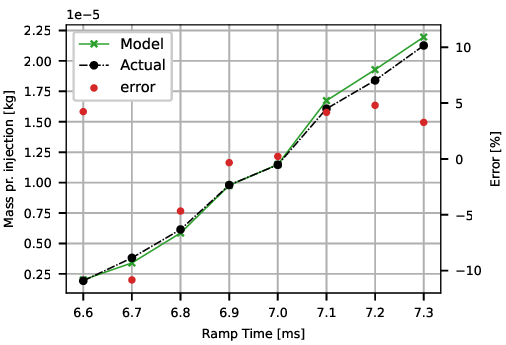}
    \caption{Dimensionless mass flow for the three tested injection pressures. }
    \label{fig:comparison}
\end{figure}

This section presents the validation and results from the experiments performed.\\
To verify the experimental setup, the ramp time is varied from 6.6 ms to 7.3 ms in 0.1 ms intervals. For each ramp time, 96 injections are saved. The 96 injections are processed as described in the Materials and Methods, and the average injected mass is found. The system is then run for 1 hour at 100 injections per minute and the oil discharged by the pressure relief valve is collected and weighed on a scale. While these measurements are taking place, several other injectors connected to the same common rail are turned on to increase the flow in the common rail and simulate a scenario closer to real operation. From the weight on the scale, a second average of the injected amount is found. The two averages are compared in Figure \ref{fig:comparison}.
As seen in Figure \ref{fig:comparison}, there is a relatively good agreement between the calculated amount from the pressure measurements and what is actually injected with the deviation being around 5\%.
\section{Conclusions}
The injection rate of a common rail swirl injection principle cylinder lubrication system was tested by the Bosch rate of injection method.
The results show:
\begin{itemize}
    \item A rise in mass flow of approximately 1 ms until cavitation in the nozzle starts choking the flow. When the choking begins, the cavitation is violent enough to atomise the oil. This results in a fast spray development, which is a strength when using the swirl injection principal for lubrication.
    \item The injected amount as a function of ramp time is approximately a straight line, making is easier to accurately dose the correct amount of oil together with the rise and fall times close to 1 ms for the mass flow.
    \item The Bosch rate of injection method is capable of predicting the delivered amount of an injection within 5\% of the weighted amount for almost the entire tested range of ramp times for the highly viscous fluid used called HydraWay HVXA15, which has similar properties to heated cylinder lubrication oil.
\end{itemize}
 Being able to accurately dose lubrication oil at from mg per injection and up, together with the fast rise and fall times for the mass flow, paves the way for new possibilities of lubrication strategies. These include being able to inject several times in a piston stroke and deliver the needed amount every time.
 
 \section{Declaration of conflicting interests}
 The author(s) declare no potential conflicts of interest with respect to the research, authorship, and/or publication of this article.
 
\section{Funding}
The research leading to these results has received funding from the Danish Innovation Foundation under grant no. 9065-00197B.
\section{ORCID iD}
\mbox{Bjørn Christian Dueholm}\hspace{2pt}\href{https://orcid.org/0000-0001-8835-4038}{\orc}\hspace{2pt}\url{https://orcid.org/0000-0001-8835-4038}\newline
\mbox{Jesper de Claville Christiansen}\hspace{2pt}\href{https://orcid.org/0000-0001-5501-7019}{\orc}\hspace{2pt}\url{https://orcid.org/0000-0001-5501-7019}\newline
\mbox{Benny Endelt}\hspace{2pt}\href{https://orcid.org/0000-0002-7013-1490}{\orc}\hspace{2pt}\url{https://orcid.org/0000-0002-7013-1490}



\begin{thebibliography}{17}
\providecommand{\natexlab}[1]{#1}
\providecommand{\url}[1]{\texttt{#1}}
\expandafter\ifx\csname urlstyle\endcsname\relax
  \providecommand{\doi}[1]{doi: #1}\else
  \providecommand{\doi}{doi: \begingroup \urlstyle{rm}\Url}\fi

\bibitem[Woodyard(2009)]{2009_WOODYARD}
Doug Woodyard.
\newblock {Introduction: A Century of Diesel Progress}.
\newblock In \emph{Pounder's Marine Diesel Engines and Gas Turbines}, pages
  ix--xxvii. Elsevier Science \& Technology, ninth edition, 2009.
\newblock ISBN 9780750689847.

\bibitem[IMO(2018)]{2018_IMO}
IMO.
\newblock {Adoption of the Initial IMO Strategy on Reduction of GHG Emissions
  From Ships and Existing Imo Activity Related To Reducing Ghg Emissions in the
  Shipping Sector}, 2018.
\newblock URL
  \url{https://unfccc.int/sites/default/files/resource/250_IMO\%20submission_Talanoa\%20Dialogue_April\%202018.pdf}.

\bibitem[Dragsted and Toft(2004)]{2004_Dragsted}
Joern Dragsted and Oyvind Toft.
\newblock {PAPER NO .: 9 Influence of low cylinder consumption on operating
  cost for 2-stroke engines}.
\newblock \emph{CIMAC Congress, Kyoto}, 2004.

\bibitem[Aabo et~al.(2001)Aabo, Liddy, Lim, and Moore]{Aabo2001}
K.~Aabo, J.~P. Liddy, K.~C. Lim, and S.~L. Moore.
\newblock {2-Stroke Crosshead Engine Cylinder Lubrication — the Future Here
  Today}.
\newblock \emph{23rd CIMAC World Congress}, pages 257--266, 2001.
\newblock \doi{10.1007/3-540-27032-9_42}.

\bibitem[He et~al.({2019a})He, Zhou, Xie, and Zhang]{2019_Yuhai}
Yuhai He, Peilin Zhou, Liangtao Xie, and Jiyun Zhang.
\newblock {New concept and design of electronically controlled cylinder
  lubrication system for large two-stroke marine diesel engines}.
\newblock \emph{{INTERNATIONAL JOURNAL OF ENGINE RESEARCH}}, {20}\penalty0
  ({8-9}):\penalty0 {967--985}, {OCT} {2019}.
\newblock \doi{10.1177/1468087418822634}.

\bibitem[He et~al.(2019b)He, Zhou, Xie, and Zhang]{He2019}
Yuhai He, Peilin Zhou, Liangtao Xie, and Jiyun Zhang.
\newblock {Design and experimental development of a new electronically
  controlled cylinder lubrication system for the large two-stroke crosshead
  diesel engines}.
\newblock \emph{International Journal of Engine Research}, 20\penalty0
  (8-9):\penalty0 986--1000, 2019.
\newblock ISSN 20413149.
\newblock \doi{10.1177/1468087418824216}.
\newblock URL \url{https://doi.org/10.1177/1468087418824216}.

\bibitem[Li et~al.(2022)Li, Jiao, Ma, Lu, and Qiao]{Li2022}
Tongyang Li, Bowen Jiao, Xuan Ma, Xiqun Lu, and Zhuhui Qiao.
\newblock {Optimization study of oil-feed parameters to improve the ring pack
  lubrication performance for a two-stroke marine diesel engine}.
\newblock \emph{Engineering Failure Analysis}, 137\penalty0 (December
  2021):\penalty0 106234, 2022.
\newblock ISSN 13506307.
\newblock \doi{10.1016/j.engfailanal.2022.106234}.
\newblock URL \url{https://doi.org/10.1016/j.engfailanal.2022.106234}.

\bibitem[Li et~al.(2021)Li, Ma, Lu, Wang, Jiao, Xu, and Zou]{Li2021}
Tongyang Li, Xuan Ma, Xiqun Lu, Chuanjuan Wang, Bowen Jiao, Hanzhang Xu, and
  Dequan Zou.
\newblock {Lubrication analysis for the piston ring of a two-stroke marine
  diesel engine taking account of the oil supply}.
\newblock \emph{International Journal of Engine Research}, 22\penalty0
  (3):\penalty0 949--962, 2021.
\newblock ISSN 20413149.
\newblock \doi{10.1177/1468087419872113}.

\bibitem[Overgaard et~al.(2018)Overgaard, Klit, and V{\o}lund]{Overgaard2018}
H.~Overgaard, P.~Klit, and A.~V{\o}lund.
\newblock {Investigation of different piston ring curvatures on lubricant
  transport along cylinder liner in large two-stroke marine diesel engines}.
\newblock \emph{Proceedings of the Institution of Mechanical Engineers, Part J:
  Journal of Engineering Tribology}, 232\penalty0 (1):\penalty0 85--93, 2018.
\newblock ISSN 2041305X.
\newblock \doi{10.1177/1350650117744100}.

\bibitem[Xiuyi et~al.(2022)Xiuyi, Jiao, Wang, Azam, Lu, Zou, Ma, and
  Neville]{Xiuyi2022}
Lyu Xiuyi, Bowen Jiao, Yuechang Wang, Abdullah Azam, Xiqun Lu, Dequan Zou, Xuan
  Ma, and Anne Neville.
\newblock {An improved contact model considered the effect of boundary
  lubrication regime on piston ring-liner contact for the two-stroke marine
  engines from the perspective of the Stribeck curve}.
\newblock \emph{Proceedings of the Institution of Mechanical Engineers, Part C:
  Journal of Mechanical Engineering Science}, 236\penalty0 (5):\penalty0
  2602--2616, 2022.
\newblock ISSN 20412983.
\newblock \doi{10.1177/09544062211027203}.

\bibitem[Jiao et~al.(2021)Jiao, Li, Ma, Wang, Xu, Lu, and Liu]{Jiao2021}
Bowen Jiao, Tongyang Li, Xuan Ma, Chuanjuan Wang, Hanzhang Xu, Xiqun Lu, and
  Zhigang Liu.
\newblock {Lubrication analysis of the piston ring of a two-stroke marine
  diesel engine considering thermal effects}.
\newblock \emph{Engineering Failure Analysis}, 129\penalty0 (March):\penalty0
  105659, 2021.
\newblock ISSN 13506307.
\newblock \doi{10.1016/j.engfailanal.2021.105659}.
\newblock URL \url{https://doi.org/10.1016/j.engfailanal.2021.105659}.

\bibitem[Milanese et~al.(2021)Milanese, Iacobazzi, Stark, and
  de~Risi]{Milanese2021}
Marco Milanese, Fabrizio Iacobazzi, Matthias Stark, and Arturo de~Risi.
\newblock {Development of common rail lube oil injector for large two-stroke
  marine diesel engines}.
\newblock \emph{International Journal of Engine Research}, 2021.
\newblock ISSN 20413149.
\newblock \doi{10.1177/14680874211008005}.

\bibitem[Payri et~al.(2021)Payri, Gimeno, Mart{\'{i}}-Aldarav{\'{i}}, and
  Viera]{Payri2021}
Ra{\'{u}}l Payri, Jaime Gimeno, Pedro Mart{\'{i}}-Aldarav{\'{i}}, and Alberto
  Viera.
\newblock {Measurements of the mass allocation for multiple injection
  strategies using the rate of injection and momentum flux signals}.
\newblock \emph{International Journal of Engine Research}, 22\penalty0
  (4):\penalty0 1180--1195, 2021.
\newblock ISSN 20413149.
\newblock \doi{10.1177/1468087419894854}.

\bibitem[Bosch(1966)]{1966_Bosch}
Wilhelm Bosch.
\newblock {The fuel rate indicator: A new measuring instrument for display of
  the characteristics of individual injection}.
\newblock \emph{SAE Technical Papers}, 1966.
\newblock ISSN 26883627.
\newblock \doi{10.4271/660749}.

\bibitem[Ravendran et~al.(2017)Ravendran, Jensen, {De Claville Christiansen},
  Endelt, and Jensen]{Ravendran2017}
Rathesan Ravendran, Peter Jensen, Jesper {De Claville Christiansen}, Benny
  Endelt, and Erik~Appel Jensen.
\newblock {Rheological behaviour of lubrication oils used in two-stroke marine
  engines}.
\newblock \emph{Industrial Lubrication and Tribology}, 69\penalty0
  (5):\penalty0 750--753, 2017.
\newblock ISSN 00368792.
\newblock \doi{10.1108/ILT-03-2016-0075}.

\bibitem[Bower and Foster(1991)]{1991_Bower}
Glenn~R Bower and David~E Foster.
\newblock {A Comparison of the Bosch and Zuech Rate of Injection Meters}.
\newblock \emph{SAE TECHNICAL PAPER SERIES}, 1991.
\newblock \doi{10.4271/910724}.

\bibitem[Ravendran et~al.(2019)Ravendran, Endelt, Christiansen, Je\~nsen,
  Theile, and Najjar]{Ravendran2019}
Rathesan Ravendran, Benny Endelt, Jesper Christiansen, Peter Je\~nsen, Martin
  Theile, and Ibrahim Najjar.
\newblock Coupling method for internal nozzle flow and the spray formation for
  v\ iscous liquids.
\newblock \emph{International Journal of Computational Methods and Experimental
  Meas\ urements}, 7:\penalty0 130--141, 03 2019.
\newblock \doi{10.2495/CMEM-V7-N2-130-141}.

\end{thebibliography}





\end{document}